\documentclass[preprint]{aastex}

\received{2001 23 May}
\accepted{2001 20 Sep}

\slugcomment{}

\newcommand{\msun}{$M_\odot$~}

\begin{document}
\title{A POSSIBLE X-RAY PERIODICITY AT SEVERAL TENS HOURS
OF A ULTRA-LUMINOUS COMPACT X-RAY SOURCE IN IC~342}

\author{
M. Sugiho$^{1}$, J. Kotoku$^{1}$, K. Makishima$^{1}$,
A. Kubota$^{2}$, T. Mizuno$^{3}$, Y. Fukazawa$^{3}$, and M. Tashiro$^{4}$}

\affil{1:Department of Physics,  University of Tokyo,
7-3-1 Hongo, Bunkyo-ku, Tokyo, Japan 113-0033}
\email{sugiho@amalthea.phys.s.u-tokyo.ac.jp}
\affil{2:The Institute for Space and Astronautical Science,
3-1-1 Yoshinodai, Sagamihara, Kanagawa, Japan 229-8510}
\affil{3:Department of Physical Science, Hiroshima University,
1-3-1 Kagamiyama, Higashi-Hiroshima, Hiroshima, Japan 739-8526}
\affil{4:Department of Physics, University of Saitama,
255, Shimo-Okubo, Saitama, Saitama, Japan 338-8570}

\begin{abstract}
A long (155 hours) {\it ASCA} observation was performed
of two ultra luminous compact X-ray sources,
Source~1 and Source~2,
in the spiral galaxy IC~342.
While Source~1 that was in a hard spectral state varied little,
Source~2 that was in its soft spectral state varied
significantly on a time scale of about one day.
The rms variation amplitude amounts to 5\% in the 2--10 keV band,
but is less than 4\% in 0.7--2 keV.
The variation involves statistically significant changes in the
parameters describing multi-color disk blackbody emission from this source.
The variation is possibly periodic,
with a period of either $31 \pm 2$ hours or $41 \pm 3$ hours.
Both are consistent with the orbital period of a semi-detached
binary formed by a black hole
and a main-sequence star of several tens solar masses.
These results reinforce the interpretation of these
X-ray objects in terms of accreting massive stellar black holes.
\end{abstract}


\section{Introduction}

Ultra Luminous compact X-ray sources (ULXs; Makishima et al. 2000)
are objects often seen in off-center regions of nearby spiral galaxies,
and exhibit X-ray luminosities by far exceeding
the Eddington limit for a neutron star
(Fabbiano. 1989; Read et al. 1997; Lira et al. 2000; Roberts \& Warwick 2000;
Bauer et al. 2001; Fabbiano et al. 2001; Strickland et al. 2001).
As reported by many {\it ASCA} observers (Colbert \& Mushotzky 1999;
Kotoku et al. 2000; Makishima et al. 2000 and references therein),
the spectra of a majority of ULXs are well fitted
by multi-color disk blackbody model
(MCD model; Mitsuda et al. 1984; Makishima et al. 1986; Ebisawa et al. 1993).
Further considering their high luminosities ($10^{39-40}~{\rm ergs~s^{-1}}$),
ULXs are hence inferred to be black-hole binaries (BHBs)
of 10--100 $M_{\odot}$
(Makishima et al. 2000; Mizuno 2000; Kubota et al. 2001).
A possible formation scenario of such massive ($\sim 100~M_{\odot}$) BHs
has been proposed by Ebisuzaki et al. (2001).
Alternatively, ULXs may be more ordinary BH (or even neutron star)
binaries with mild X-ray beaming (King et al. 2001),
although realistic X-ray beaming mechanisms are yet to be worked out.

Through two {\it ASCA} observations of the spiral galaxy IC~342,
Kubota et al. (2001; hereafter Paper~I) discovered that the two ULXs in it,
called Source~1 and Source~2 (Fabbiano \& Trinchieri 1987),
make clear transitions between soft and hard spectral states.
Since such a transition is
characteristic of Galactic/Magellanic BHBs
(e.g., Maejima et al. 1984; Zhang et al. 1997; Wilms et al. 2000),
the BHB interpretation of ULXs has been much reinforced.
A similar spectral transition was observed from M81 X-9
(La Parola et al. 2001), thought to be a ULX as well.

To conclusively establish the binary nature of ULXs,
we in fact need one more step,
i.e., to detect their binary periods.
Although the standard optical technique
(Gies \& Bolton 1982; Cowley et al. 1983;
for review Cowley et al. 1992)
is currently unapplicable to ULXs due to
the general lack of their optical identification,
we may search their X-ray light curves for possible periodicity.
According to the Roche-lobe formula (Eggleton 1983)
for a semi-detached binary system and Kepler's law,
the density $\rho$ (g cm$^{-3}$) of
the mass-donating star filling its Roche-lobe
is determined by the binary period $P$ (hour) as $\rho \simeq 115 P^{-2}$,
if the mass-accreting star is more massive than the mass-donating star
(Frank et al. 1992).
For example, if a ULX indeed consists of a BH of $100~M_{\odot}$
together with a $50~M_{\odot}$ main-sequence companion,
we expect $P \sim 40$ hours,
because the latter has typically a radius of $\sim$ 10 $R_{\odot}$
(i.e., $\rho \sim 0.07$),
although it may finally evolve to a much larger size
(De Loore et al. 1978; Weiss 1994; Stothers \& Chin 1999).
Actually, in NGC~5204,
a possible optical counterpart to a ULX has been discovered,
and may be considered to be a supergiant O star (Roberts et al. 2001).
If the companion mass is instead $10~M_{\odot}$ ($\sim$ 4 $R_{\odot}$ radius),
we expect $P \sim 20$ hours.
Thus, the orbital periods of such binary systems
are expected in the range of several tens of hours.

When the first {\it ASCA} observation of IC~342 was conducted in 1993,
a significant X-ray
variation was detected from Source~1
on the time scale of $\sim$ 10~hours (Okada et al. 1998).
However, the total span of that observation
($\sim$ 22~hours) was too short
to examine whether the variation is periodic or not.
In this paper, we report timing results on the two ULXs in IC~342,
Source~1 and Source~2,
based on the one-week {\it ASCA} observation conducted in 2000.
We have indeed obtained evidence for $\sim$ 31 or $\sim$ 41~hour
periodicity from Source~2.

\section{Observation and Data Reduction}

The present {\it ASCA} observation of IC~342,
the second one described in Paper~I, was carried out
from 2000 February 24 00:29 through March 1 21:09 (UT),
for a total time span of 557~ks.
After screening the data with the standard criteria,
we obtained the net exposure of 262~ks with the GIS
and 244~ks with the SIS.
As already described in Paper~I,
this observation found Source~1 in the hard spectral state
while Source~2 in the soft spectral state.
Referring to the X-ray images (Paper~I),
we accumulated photons over circular regions of $3'$ radius
around the two sources.

\section{Results}


\subsection{Power-density Spectra}

\begin{figure}
\epsscale{0.9}
\plotone{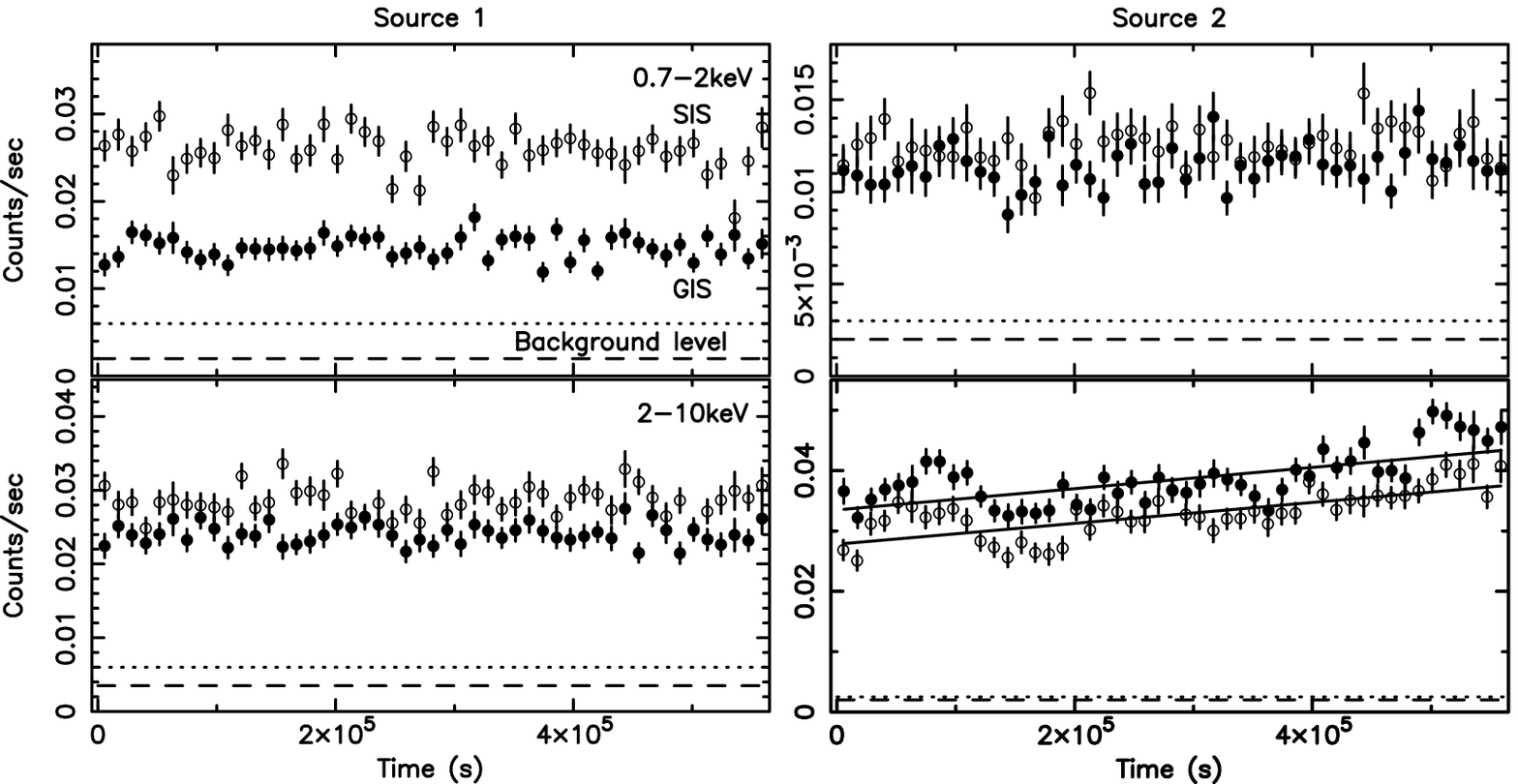}
\figcaption{
The GIS (filled circles) and SIS (open circles) light curves of
Source~1 (left column) and Source~2 (right column),
in the energy band of 0.7--2~keV (upper panels) and 2--10~keV (lower panels).
Background rates are indicated by dashed lines for the GIS and by dotted
lines for the SIS.
The data bin width is 11.5 ks, corresponding to twice the orbital period of
the spacecraft.
Two solid lines in the bottom right panel indicate the best-fit linear trend.
\label{fig:lightcurves}}
\end{figure}

Figure \ref{fig:lightcurves} shows the
GIS and SIS two-band light curves obtained from
the two sources.
The photon arrival times have been converted to solar barycentric times.
The background counts contained in the light curves were evaluated
using an off-source region in the image for the GIS and 
other blank-sky observations for the SIS, respectively,
because no off-source region is available in the SIS image.
Thus, the background contributes $\sim 10 \%$ to the light curves.
Furthermore, the GIS background is stable within $\sim 10 \%$ (rms, relative),
implying that the background variation is below $1 \%$ of the GIS signal
counts.
The variation of SIS background is known to be similar to that of the GIS.
Thus, the background and its variation can be ignored in our subsequent 
timing study.
In figure \ref{fig:lightcurves},
we can clearly see significant variations on $\sim$ one day time scale
in the high-energy light curves of Source~2,
consistently with the GIS and SIS data.
In contrast, the other light curves appear to be consistent with being constant.
Actually, the value of reduced $\chi^2$ obtained by fitting a linear trend,
indicated in Figure \ref{fig:lightcurves},
is 2.9 and 2.3 for the GIS and SIS high-energy light curves of Source~2,
respectively, while it is below 1.5 for the other light curves.

\placefigure{fig:lightcurves}

\begin{figure}
\epsscale{0.45}
\plotone{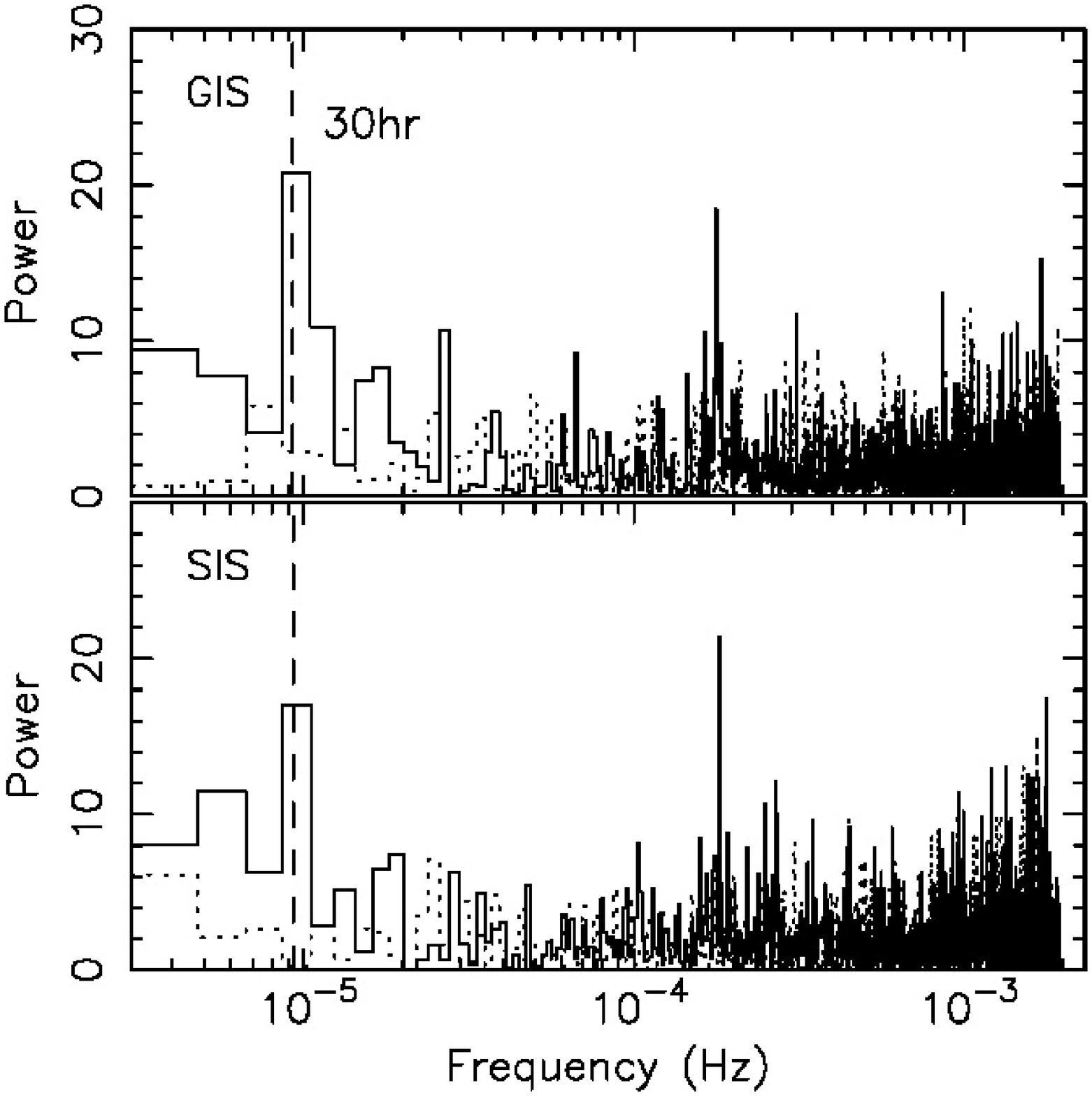}{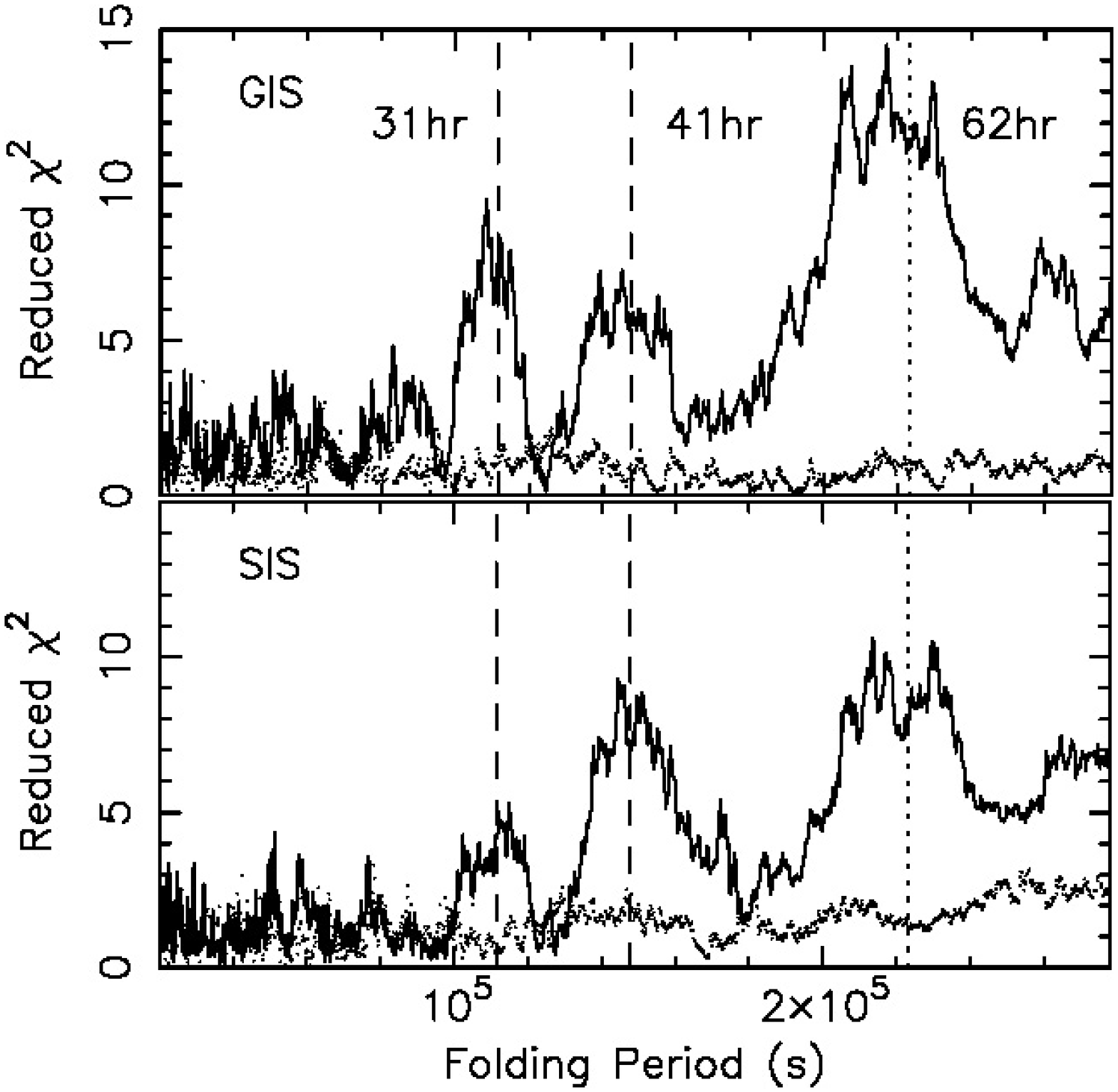}
\figcaption{
The power density spectra of the 2--10~keV light curves of Source~2
(solid lines) and of Source~1 (dotted lines)
from the GIS (top panel) and from the SIS (bottom panel).
The dashed vertical line indicates a period of 30 hours.
\label{fig:power_spectra}}
\end{figure}

To examine the 2--10~keV GIS and SIS light curves of
Source~2 for the periodic behavior,
we re-calculated them with a finer bin width of 256~s,
and embedded them into a time span of 524~ks = 146~h (2048 bins).
We next subtracted the linear trend from the light curves
and filled the data gaps with zeros.
Using an FFT algorithm,
we calculated power density spectra,
separately for the two instruments, over a frequency range from
$3.8\times 10^{-5}$~Hz ($2^{18}$~s $=262$~ks $=73$~hours) to
$2.0 \times 10^{-2}$~Hz ($2^{9}=512$~s).
Figure \ref{fig:power_spectra} shows the power density spectra of
the 2--10~keV GIS and SIS light curves of Source~2,
together with that of Source~1 in the same energy band.
We thus detect a strong power peak at
$\sim 9 \times 10^{-6}$~Hz (period $110$~ks $=30$~hours) from Source~2,
both instruments showing consistent results.
The chance probability to detect such a strong power below $10^{-4}$~Hz
with the two instruments, assuming that the source is varying randomly
with the same rms amplitude as the actual data,
is estimated by a Monte Carlo simulation to be $\sim 5 \times 10^{-5}$.
The $\sim 9 \times 10^{-6}$~Hz peak is absent in the 2--10~keV data
of Source~1 (Figure \ref{fig:power_spectra}),
indicating that the peak is unrelated to background.
In the 0.7--2~keV band,
neither of the two sources exhibits the $\sim 9 \times 10^{-6}$~Hz peak.
Another peak at $\sim 1.8 \times 10^{-4}$~Hz ($5.6$~ks)
corresponds to the orbital period of {\it ASCA}.

\placefigure{fig:power_spectra}

\subsection{Epoch-folding analysis}

\begin{figure}
\epsscale{0.45}
\plotone{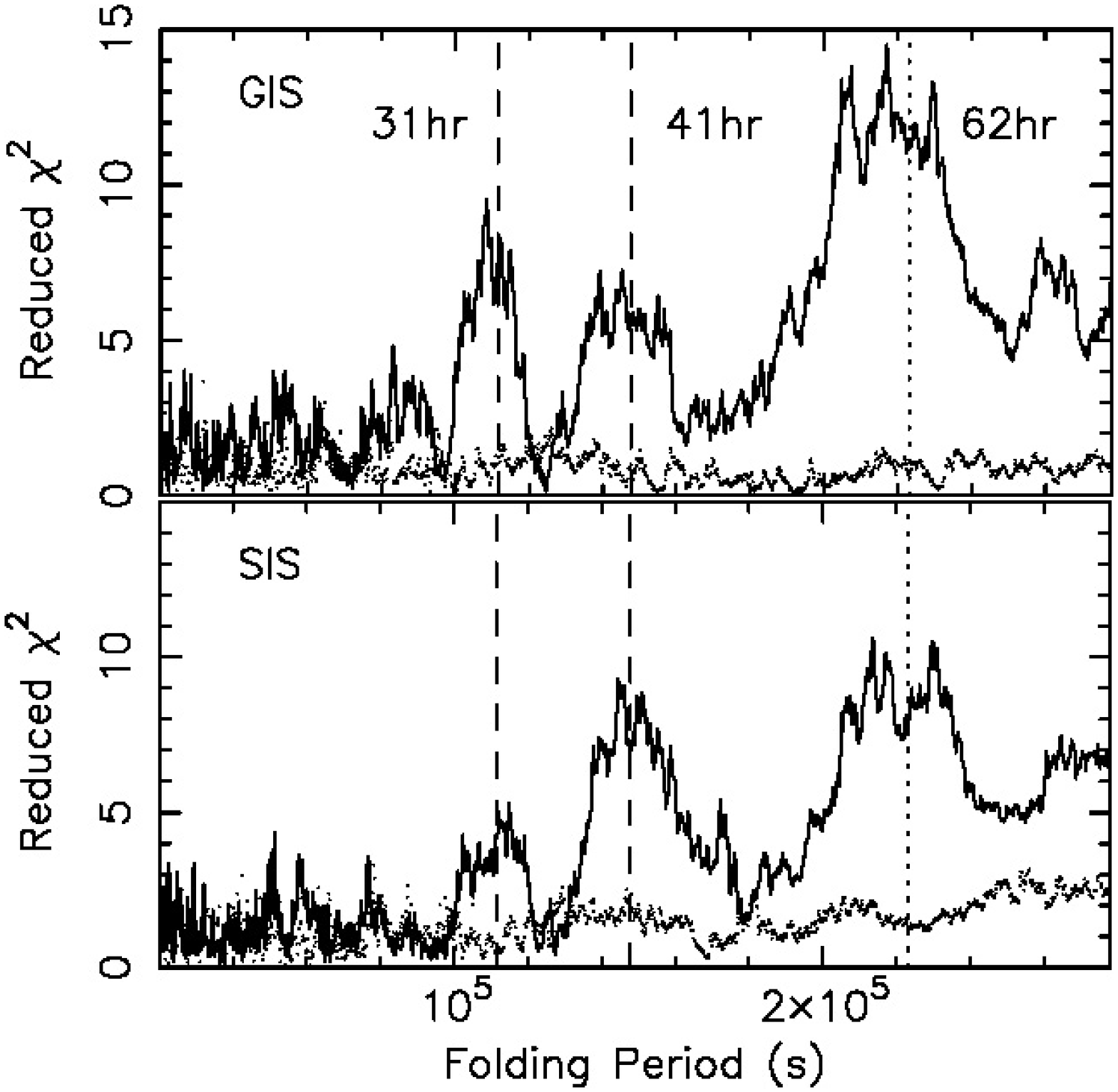}
\figcaption{
The periodogram for the 2--10~keV GIS (top panel)
and SIS (bottom panel) light curves of Source~2 (solid lines)
and of Source~1 (dots),
calculated over a period range of 20~ks to 278~ks with a step of 500~s.
The periods of significant variation (31 and 41 hours)
are indicated by dashed lines,
while the double period (62 hours) of the former by dotted lines.
\label{fig:epoch_folding}}
\end{figure}

In order to further examine the possible periodicity found in the
power-density spectra,
we performed standard epoch-folding analysis.
That is, we folded the overall light curve (of either GIS or SIS)
of Source~2 into 7 bins at various trial periods,
and examined the obtained folded light curves against the
hypothesis of ``being constant'' through $\chi^2$-evaluation.
Figure \ref{fig:epoch_folding} shows the GIS and SIS periodograms of Source~2,
thus calculated in the 2--10~keV band,
together with that of Sources~1 obtained in the same way.
In the periodograms of both instruments for Source~2,
we observe clear peaks at periods of
$\sim 110$~ks ($31 \pm 2$ hours) and $\sim 145 $~ks ($41 \pm 3$ hours).
The reduced $\chi^{2}$ values are
$8\sigma$ (GIS) and $5\sigma$ (SIS) for the 31 hour period,
and $6\sigma$ (GIS) and $9\sigma$ (SIS) for the 41 hour period;
their significances are similarly high.
These periods are not present in the 2--10~keV periodogram of Source~1
(Figure \ref{fig:epoch_folding}),
or 0.7--2~keV periodogram of  either source.
The peak periods do not depend significantly
on the number of bins used in the folding.
The double period of 31 hours is also seen at $\sim 62$ hours.
We do not find other significant peaks in the periodograms.

\placefigure{fig:epoch_folding}

The 31 hour period of Source~2
means about 5 cycles across the total data span,
and corresponds to the peak found at $9.0 \times 10^{-6}$~Hz
in the power density spectra (Figure \ref{fig:power_spectra}).
The 41 hour period, in contrast, corresponds to about 3.8 overall cycles,
and its counterpart (to be expected at $6.8 \times 10^{-6}$~Hz)
is weakly visible in the power density spectra, especially in the SIS data.

\begin{figure}
\epsscale{0.45}
\plotone{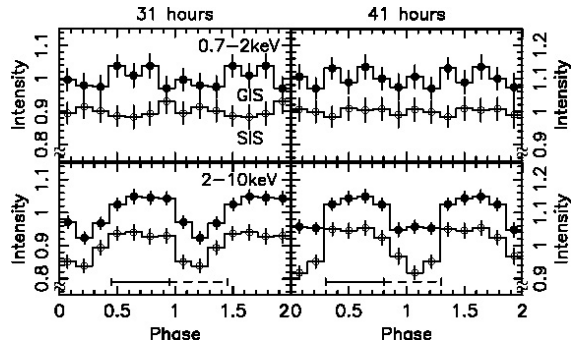}
\figcaption{The GIS (filled circles, with the scale to the left)
and SIS (open circles, with the scale to the right)
light curves of Source~2 folded with a period of
31 hours (left panels) and 41 hours (right panels),
in the 0.7--2~keV band (upper panels) and the 2--10~keV band (lower panels),
shown for two cycles.
Phase 0 corresponds to 2000 February 24 00:30:00.
Bright and faint phases are indicated by solid and dashed lines,
respectively.
\label{fig:folded_lightcurves}
}
\end{figure}

Figure \ref{fig:folded_lightcurves} shows
the GIS and SIS two-band folded light curves of Source~2,
corresponding to the above two candidate periods (31 and 41 hours).
Thus, a roughly sinusoidal modulation of $\sim $5\% amplitude
is seen in the high-energy band,
whereas the modulation is insignificant in the low-energy band
with an upper limit of 4.0\% and 3.9\% (rms) for the 31 hour and
41 hour periods, respectively.
These properties do not differ by a large amount between the two 
candidate periods.

\placefigure{fig:folded_lightcurves}

\subsection{Phase-resolved spectra}

\begin{figure}
\epsscale{0.45}
\plotone{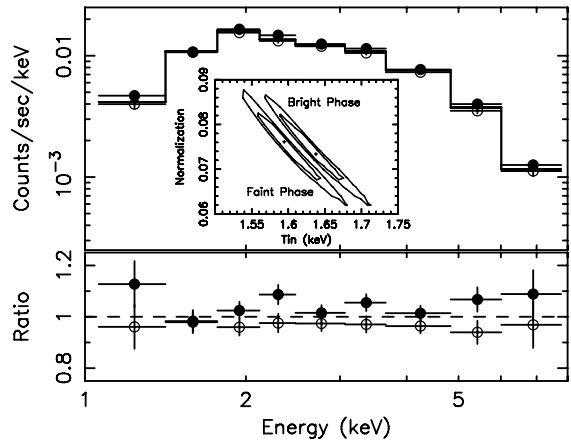}
\figcaption{
The GIS spectra of Source~2, over the bright phase
(filled circles) and faint phase (open circles)
of the 31 hour periodic variation.
The histograms show the best-fit absorbed MCD model for the time-averaged 
spectrum,
as obtained in Paper~I.
The lower panels indicate their ratios to the best-fit model.
The inset shows the 90~\% and 68~\% confidence contours in the plane of
disk temperature and normalization.
\label{fig:spectra}}
\end{figure}

To investigate spectral changes associated with these periods,
we sorted the GIS spectra into two phases (bright and faint)
of the modulation,
as indicated in Figure \ref{fig:folded_lightcurves},
separately for the two possible periods.
We did not analyze the SIS spectra,
since the long-term CCD degradation has made its response uncertain.
The obtained phase-resolved GIS spectra for the 31 hour period
are presented in Figure \ref{fig:spectra},
together with ratios to the MCD model
that best fits the phase-averaged GIS spectrum (Paper I).
Thus, the spectrum changes by $\sim \pm 10$\%,
with the change being larger in the harder band.

Similarly to Paper I, we fitted the two
phase-resolved GIS spectra successfully with the MCD model.
Because the insignificant modulation in lower energies
means a constant absorption across the modulation phase,
we fixed the absorbing column to $N_{\rm H} = 1.8 \times 10^{22}$ cm$^{-2}$
which was found through the phase-averaged spectral analysis (Paper I).
The inset in Figure \ref{fig:spectra}
shows the obtained confidence contours on the plane of the two MCD model 
parameters,
disk temperature and normalization.
Thus, the two confidence regions do not overlap with each other,
implying that the periodic modulation involves statistically significant 
changes
in the accretion disk condition (the temperature and/or the apparent radius).
Although the linear trend shown in Figure \ref{fig:lightcurves} has a 
similar effect,
it contributes little ($<$ 1\%) to this spectral modulation
because the linear trend is averaged out by the folding process.
Although Figure \ref{fig:spectra} refers to the 31~hour period,
the 41~hour period yields very similar results.

\section{Discussion}

During the {\it ASCA} long-look observation of IC~342 conducted in 2000,
Source~2, in its soft spectral state, varied clearly.
The variation has a typical time scale of one day,
is more prominent in higher energies,
and can be understood as changes in the MCD temperature and/or normalization.
These properties are reminiscent of the variation detected
in the shorter {\it ASCA} observation in 1993 from Source~1,
when it was in the high state (Okada et al. 1998).
Mizuno (2000) and Mizuno et al. (2001) observed similar variation
from several other ULXs in the spectral high state.
Thus, a mild intensity variation may be a relatively
common phenomenon among high-state ULXs.

Our timing analysis indicates
that the Source~2 variation is possibly periodic,
with a suggested period of either 31 hours or 41 hours.
Of course, we cannot conclude for sure
on the cyclic nature of the variation,
nor can we distinguish between the two candidate periods,
because the time span of the observation covers
only 4--5 cycles of the suggested periods.
Nevertheless, these values agree with what is expected
for the orbital period of a semi-detached binary system
consisting of a BH and a massive main-sequence star.
Specifically, the $P$ vs $\rho$ relation described in \S~1
yields a main-sequence star mass of
$\sim$ 25 \msun and $\sim$ 50 \msun for $P=$ 31 and 41~hours, respectively;
a lower mass is indicated, if the mass-donating star is evolved and hence
has a larger size.
However, the BH mass remains rather unconstrained.

If the observed variability of Source~2 really reflects an orbital effect,
the phase-dependent spectral changes may be caused, e.g.,
by an orbital modulation in the mass accreting rate,
or eclipses of the inner disk.
Actually, based on a 6.4-day observation with {\it RXTE},
Boyd et al. (2001) report on a clear high-energy X-ray modulation
in LMC~X-3 synchronized with its 1.7-day orbital period.

Recently, Bauer et al. (2001) detected a very clear 7.5-hour periodic
X-ray variation from a ULX in the Circinus Galaxy.
If the period is orbital in origin,
the implied system has a few \msun main-sequence star and a $\sim 30$
\msun BH,
assuming that its X-ray luminosity ($ 3.6\times 10^{39}~{\rm ergs~s^{-1}}$)
is close to the Eddington limit without beaming.
This source may be a scale-down version of Source~2 in IC~342.

In conclusion,
the present results significantly reinforce the interpretation of ULXs
as mass-accreting BH binaries.


\clearpage



\end{document}